\documentclass[12pt,preprint]{aastex}

\shorttitle{Grain Retention in MRI Turbulent Disks}
\shortauthors{Kretke and Lin}

\begin{document}
\title{Grain Retention and Formation of Planetesimals near the 
Snow Line in MRI-driven Turbulent Protoplanetary Disks}
\author{Katherine A. Kretke$^1$ and D. N. C. Lin$^{1,2}$}
\affil{$^1$Department of Astronomy and Astrophysics, University 
of California, Santa Cruz, USA}
\affil{$^2$Kavli Institute of Astronomy and Astrophysics, Peking 
University, Beijing, China}
\email{kretke@ucolick.org}

\begin{abstract}
The first challenge in the formation of both terrestrial planets and the cores of gas giants is the retention of grains in protoplanetary disks. In most regions of these disks, gas attains sub-Keplerian speeds as a consequence of a negative pressure gradient. Hydrodynamic drag leads to orbital decay and depletion of the solid material in the disk, with characteristic timescales as short as only a few hundred years for meter-sized objects at 1 AU.  In this paper, we suggest a particle retention mechanism which promotes the accumulation of grains and the formation of planetesimals near the water sublimation front or ``snow line.'' This model is based on the assumption that, in the regions most interesting for planet formation, the viscous evolution of the disk is due to turbulence driven by the magneto-rotational instability (MRI) in the surface layers of the disk. The depth to which MRI effectively generates turbulence is a strong function of the grain size and abundance. A sharp increase in the grain-to-gas density ratio across the snow line reduces the column depth of the active layer. As the disk evolves towards a quasi-steady-state, this change in the active layer creates a local maximum in radial distribution of the gas surface density and pressure, causing the gas to rotate at super-Keplerian speed and halting the inward migration of grains. This senario presents a robust process for grain retention which may aid in the formation of proto-gas-giant cores preferentially near the snow line.
\end{abstract}

\section{Introduction}
More than 200 extrasolar planets have been discovered with radial
velocity surveys, corresponding to 10 to 15\% of the sample of mostly nearby 
solar-type stars(Marcy {\it et al.} 2005).
In the current paradigm, formation of both terrestrial and gas giant planets
begin with the collisional growth of planetesimals, the latter only
accreting gas once the solid cores have reached several earth masses 
(Pollack {\it et al.} 1996).  The growth rate of these protoplanetary 
embryos is determined by the surface density 
distribution of planetesimals, $\Sigma_p$, in their nascent disks.  
In the typical construction of planet-formation sequence (eg. Kokubo \& Ida 
2002, Ida \& Lin 2004, hereafter IL04), authors have often utilized an
empirical ``minimum mass nebula model'' (MMNM) in which $\Sigma_p \simeq 
10 r_{\rm AU}^{-3/2}$g~cm$^{-2}$, where $r_{\rm AU}$ is the radial 
distance from the central star. This {\it ansatz}
is based on the assumption that the planets in the solar system
locally retained all the heavy elements and a fraction of the gas
in the primordial solar nebula (Hayashi {\it et al.} 1985). 
Within the observationally inferred range of protostellar 
disk masses (Beckwith \& Sargent 1996), gas giants can readily form near
the snow line on disks with 2 to 10 times enhancement over MMNM(IL04).

Despite the successful reproduction of the planets' observed mass-period 
distribution, the global scaling of $\Sigma_p$ with that of the MMNM 
is questionable. The main uncertainties are: (1) whether heavy elements can be 
efficiently incorporated into planetesimals which do not migrate significantly
and (2) whether there is a sufficient $\Sigma_p$ enhancement near the snow-line 
to favor the emergence of gas giant planets. The observed surface 
density of the dust, $\Sigma_d$, is in the form of grains ranging 
from a few $\mu$m to mm. These entities must grow into embryos larger 
than $10^3$~km in gaseous disks.  In most regions of the disk, 
the pressure, $P$, decreases with $r$ 
so that the azimuthal velocity of the gas, $V_\phi$, is sub-Keplerian. Grains larger 
than a few cm are decoupled from the gas and move at Keplarian speeds so they experience 
head winds and undergo orbital decay on time scales smaller than 
the growth time scale by a factor $Z=\Sigma_d /\Sigma_g$ 
(Garaud {\it et al.} 2004). In principle, grains at a few AU can be retained and then
converted into planetesimals either through sedimentation followed by 
gravitational instability (Goldreich \& Ward 1973; Youdin \& Shu 2002; 
Garaud \& Lin 2004) or through cohesive collisions (Supulver \& Lin 2000)
after the photoevaporation of disk gas increases $Z$ above unity.  However, the 
severe depletion of the disk gas will also limit the supply for 
the gas giants to accrete their massive envelopes.  

In this letter, we propose a grain retention mechanism which can promote rapid planet formation around sublimation fronts in protoplanetary disks. We assume 
the disk evolution during the classical T Tauri phase to be regulated by 
a magneto-rotational instability (MRI; Balbus \& Hawley 1991).  However, MRI 
only operates in regions with sufficient ionization
fraction to couple the gas to the magnetic field. The inner regions 
of typical protostellar disk are thermally ionized (Pneuman and 
Mitchell 1965; Umebayashi 1983).  At greater distances from the disk 
center ($r > 0.1$ AU), stellar x-rays and diffuse cosmic rays ionize 
the surface layers of the disk down to a certain column density, $\Sigma_A$
(Glassgold {\it et al.} 1997).  Thus, the disk may have a layered structure in which
the MRI-induced accretion flow occurs on its active surface which surrounds
a highly neutral and inactive ``dead zone'' (Gammie 1996). The magnitude
of $\Sigma_A$ depends strongly on the relative abundance and size 
distribution of grains (Wardle \& Ng 1999;  Sano {\it et al.} 2000).  Across 
the snow line or the dust destruction front, there are significant 
changes in the amount of solid material in the form of grains. In 
a disk with a solar composition, the total mass fraction in grains 
decreases by at least 50\% due to the sublimation of water ice 
interior to the snow line (Lodders 2003). Efficient inward 
transport of solid materials combined with the slow diffusion of the heavy elemental 
vapor can further enhance
the contrast in grains' surface density (Cuzzi \& Zahnle 2004).
This rapid radial change in dust properties and the corresponding 
rapid change in the column density in the active region leads to local maxima in  
$\Sigma_g$ and $P$ in a steady-state disk. As noted by other authors (Bryden {\it et al.} 1999; 
Haghighipour \& Boss 2003), this pressure-gradient reversal induces 
a super-Keplerian $V_\phi$ and causes solid material to build up near 
the local pressure maxima.  

In order to highlight the dominant process  
we adopt a simple, 1-d, steady-state model for the gas distribution, 
in which the gas mass flux in the radial direction, $\dot M_g = 2 \pi 
\Sigma_g V_r r$ (where $V_r$ is the radial velocity of the gas), is 
independent of $r$.   Interior to 5 AU in a typical protoplanetary disk the disk should rapidly relax to this quasi-steady-state (Beckwith \& Sargent 1996).  In \S 2, we present a simple analytical model for 
the structure of a disk with variable viscosity and describe the 
conditions under which these changes may induce a pressure-gradient
reversal. In \S3, we present numerical results to show how these 
changes manifest themselves around the dust destruction front and 
the snow line, at different stages in the disk evolution.  
Finally, we summarize our results and discuss their implications in \S4.

\section{Model Description}
We will parametrize our model in the framework of the familiar {\it ad hoc} 
$\alpha$-prescription in which the efficiency of angular momentum transport is 
approximated by an effective viscosity $\nu = \alpha(r) c_s h$, 
where $c_s$ and $h=c_s/\Omega_K$ refer to the sound speed at 
the mid plane and the isothermal density scale height, respectively (Shakura 
\& Sunyaev 1973).  
However, we deviate from the standard model in that we allow $\alpha$ to vary with $r$.  
The Keplerian frequency around a star with mass $M_\ast$
is  $\Omega_K =(G M_\ast /r^3)^{1/2}$.  
The surface density at each radius in a disk with mass accretion rate $\dot{M}_g$ is $\Sigma_g = \dot{M}_g/(3\pi\nu(r))$ (Pringle 1981).

The interaction of the gas with the sedimented grains (super-cm  
size) leads to their orbital evolution. The speed of their migration 
depends on their sizes, and the radial velocity 
of the maximally affected particles is
\begin{equation}
	v_r = \frac{1}{2\rho}\frac{\partial P}{\partial r}\Omega_K^{-1},\label{veleq}
\end{equation}	
(see Weidenschilling (1977) for derivation).
In most regions of the disk $P$ monotonically 
decreases with $r$ so that $V_\phi < \Omega_K r$ and gas drag causes the 
grains to undergo orbital decay.  However, if $\partial P/\partial r> 0$ then $v_r > 0$, and particles will drift outwards rather than inwards.  If we assume a vertically isothermal disk, the pressure at the midplane is given by $P=\Sigma_g c_s^2/(2\pi h)$.  If we assume a power-law temperature distribution in the radial direction, $T=T_0 r_{\rm AU}^{-q}$, and use the steady-state $\Sigma_g$, then $v_r$ from eq(\ref{veleq}) will be positive if
\begin{equation}
	\frac{r}{\alpha}\frac{d \alpha}{dr} < -\left(3-\frac{q}{2}\right).
\end{equation}
We can approximate a rapid radial variation in the viscosity at radius $r_0$ as a smoothed step function in $\alpha$,
\begin{equation}
	         \alpha(r) = \frac{(\alpha_1-\alpha_2)}{2}\left[1+{\rm erf}
		 \left(\frac{r_0-r}{\Delta r}\right)\right]+\alpha_2, \label{alpha}
\end{equation}
where $\Delta r$ is the characteristic width of the transition zone.
For this simple $\alpha(r)$ prescription, 
the particle outflow condition is
\begin{equation}
	\frac{\Delta r}{r} < \sqrt{\frac{1}{\pi}} 
\frac{\Delta \alpha}{\alpha} \frac{1}{(3-q/2)} \label{c_eq},
\end{equation}
where $\Delta \alpha = \alpha_1 -\alpha_2$ and $\alpha = (\alpha_1 
+ \alpha_2)/2$.

In a layer-accreting disk there is physical motivation to describe the variations in $\alpha$ with respect to the depth of the region unstable to MRI turbulence.  In this case $\alpha$ is the vertically integrated value
including both the efficient angular momentum transport from 
the active layers in the disk and a small residual contribution $\alpha_0$
from the MRI stable region such that 
\begin{equation}
\alpha=\frac{\Sigma_A}{\Sigma} \alpha_{\rm MRI}+ \alpha_0, \label{effvisc}
\end{equation}	
where $\alpha_{\rm MRI}=1.8\times 10^{-2} (\beta/1000)^{-1} $
is the appropriate scaling for MRI turbulence (Sano {\it et al.} 1998).  
For the plasma parameter ($\beta=P/P_{\rm mag}$) we use $\beta=1000$. For disks with weak magnetic fiends, reasonable values for $\beta$ are on the order of 100 to 1000 (Sano {\it et al.} 2000). 

Although an actual detailed description of $\alpha_0$ is not important 
for the effect we consider here, it represents  
some mechanism of angular momentum transport in
the ``quiescent'' part of the disk beyond the dust-destruction front (e.g. 
convective instability in the vertical direction (Lin \& Papaloizaou 1980), 
gravitational torques (Laughlin \& Rozycza 1996), linear Rossby wave 
instability (Li {\it et al.} 2000), baroclinic instability (Klahr \& 
Bodenheimer 2003; Klahr 2004),  linear stratorotational instability 
(Dubrulle {\it et al.} 2005; Shalybkov \& R\"udiger 2005), etc.)
so that it is possible for the disk gas to evolve
into a quasi-steady-state, in which $\dot M_g$ becomes an essentially
independent function of $r$.   In the absence of any other 
angular momentum transport mechanism ({\it i.e.} $\alpha_0=0$), a 
steady state is unattainable and gas will continue to accumulate  until another 
instability is generated.  A pressure-gradient 
inversion can still develop, albeit it will be 
non-trivial to retain the accumulated grains if the disk becomes
gravitationally unstable.  

The steady-state surface density profile is:
\begin{equation}
    \Sigma_g = \frac{\dot{M}}{3\pi\alpha_0 c_s h} 
- \Sigma_A\frac{\alpha_{\rm MRI}}{\alpha_0}.
\end{equation}
The column depth of the region unstable to MRI turbulence can locally be described as a powerlaw $\Sigma_A~\propto~r^p$ when the total column depth $\Sigma~>~\Sigma_A$. 
At a sublimation front in the layer-accreting region of the disk, the active region may be described locally as
\begin{equation}
	\Sigma_A = \left(\frac{r}{r_0}\right)^p \left[\frac{(\Sigma_{A,1}-\Sigma_{A,2})}{2}\left[1+{\rm erf}\left(\frac{r_0-r}{\Delta r}\right)\right]+\Sigma_{A,2}\right].
\end{equation}
$\Sigma_{A,1}$ and $\Sigma_{A,2}$ are the column depth at that radius assuming the the two different particle populations on either side of the transition.  
In this case the particle outflow condition is
\begin{equation}
	\frac{\Delta r}{r_0} < \sqrt{\frac{1}{\pi}}\frac{\Delta \Sigma_A}{\Sigma_A}\left[\frac{\Sigma_{\rm MRI}}{\Sigma_A}\left(3 -\frac{q}{2}\right) + p - 3 - \frac{q}{2}\right]^{-1},
\end{equation}
where $\Sigma_{\rm MRI} = \dot{M}_g/(3 \pi \alpha_{\rm MRI} c_s h)$ is the steady-state surface density if the entire disk were MRI active.  
In the case when $\Sigma_{\rm MRI} \gg \Sigma_A$ this solution reduces to 
\begin{equation}
	\frac{\Delta r}{r_0} < \sqrt{\frac{2}{\pi}}\frac{\Delta \Sigma_A}{\Sigma_{\rm MRI}}\frac{1}{(3-q/2)}.
\end{equation}
Note that this is independent of $\alpha_0$, the magnitude of the residual viscosity.  However, it is only meaningful in situations where $\alpha_0$ is large enough that the disk can reach a steady-state solution without becoming unstable.

\section{Results}
\subsection{Dust Destruction Front}
The modified step function in eq(\ref{alpha}) can be most easily 
applied in the inner region of the disk near the dust destruction front.  
The gas is thermally ionized 
until $T\sim 1000$K, so the 
dust-destruction front is roughly coincident with the outer boundary of
the thermally ionized region.  At this location, the disk makes a
transition from being completely MRI-active to having a very large
gas column density which is stable to MRI\@.  As the active region 
switches from covering the entire thickness of the disk to being virtually 
non-existent in the presence of a significant population of small 
particles (Turner {\it et al.} 2007), we can utilize our simple 
$\alpha$ description from eq(\ref{alpha}) where $\alpha_1 
= \alpha_{\rm MRI}$ and $\alpha_2 =\alpha_0$.  

\begin{figure}
\plotone{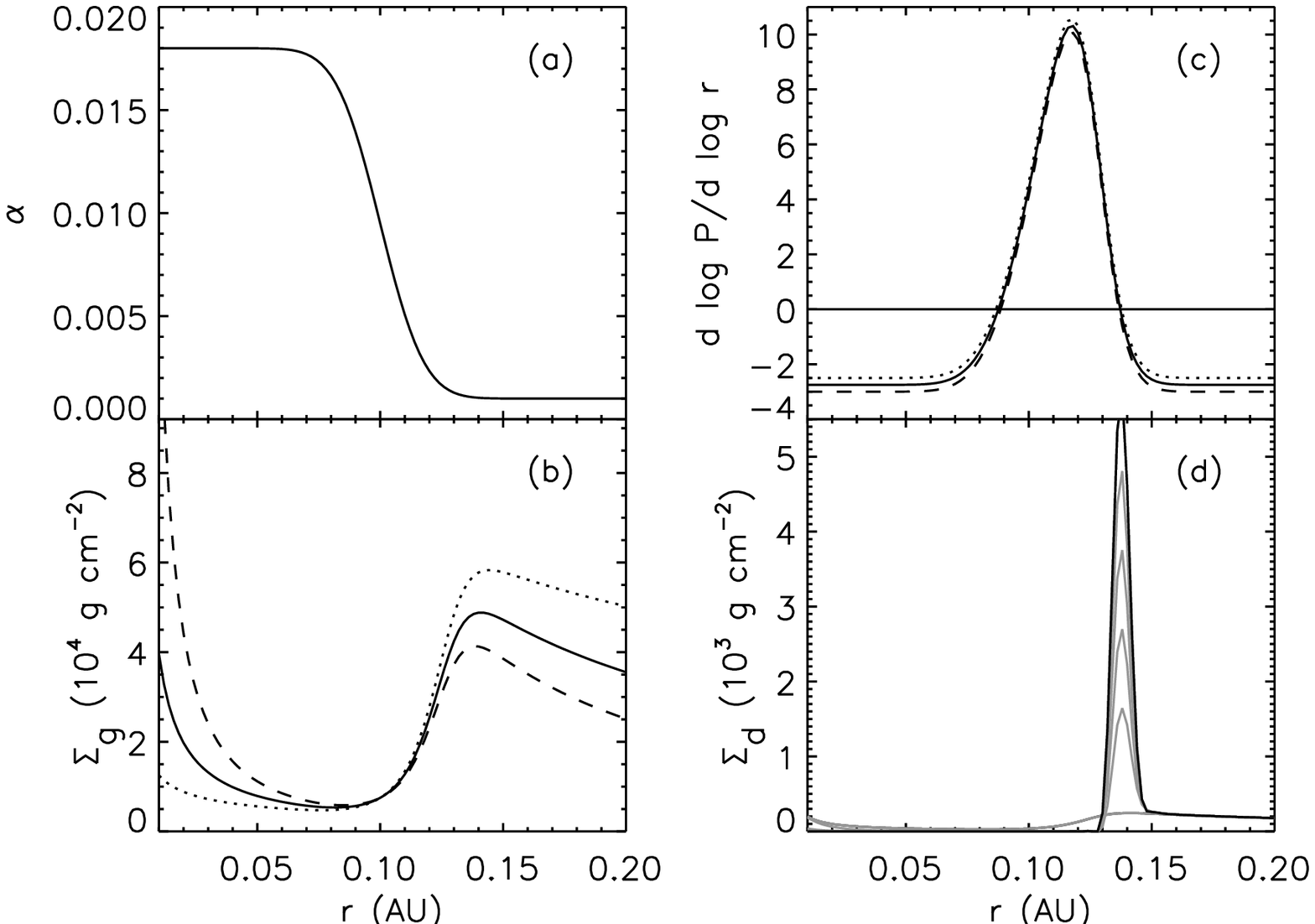}
\caption{The effect of an example variable $\alpha(r)$ shown in panel (a) on the gas properties.  Panel (b) shows the corresponding steady-state gas distribution with temperature profile $T=T_0r_{AU}^{-q}$. The curves represent $q$=0.25, 0.5, and 0.75 (dotted, solid and dashed, respectively).  Panel (c) shows the corresponding pressure gradient.  Panel (d) shows the evolution of an initially well-mixed solid population after 500 orbits at 0.1 AU (grey lines show every 100 orbits). \label{fig1}}
\end{figure}

Due to the expected large decrease in viscosity, the condition 
stated in eq(\ref{c_eq}) is easily satisfied at the 
dust-destruction front for even rather broad transition regions.
Therefore, we use this clean case to demonstrate the manner in which 
the pressure gradient is affected by the viscosity in a steady-state disk.
Fig.~\ref{fig1} shows the effect for a particular $\alpha$ profile 
with $\alpha_0 = 10^{-3}$ for different power-law temperature profiles.
While the surface density changes according to 
the temperature profile, the pressure gradient is very weakly dependent 
on temperature. 
We also show the evolution of an initially well-mixed population of maximally migrating particles.  While a realistic population of particles will not accumulate as quickly as this maximally drifting population, this is illustrative of the efficiency of particle retention at the pressure maxima.
It is also important to note that the location of the pressure maximum is exterior to the 
sublimation front, meaning that the material trapped in this region will not be lost due to sublimation.

\subsection{Snow Line}
A reversal of the pressure gradient around the snow line will only occur at certain stages in the disk evolution.  In an MRI-driven disk the structure of the disk at the snow line can be described in three stages. Initially, the column density at the snow line is very large, corresponding to a high steady-state mass accretion rate.   The MRI-active layer is negligible due to the presence of numerous small grains.  If the mass accretion rate is high 
(as in disks around FU Ori systems), $\Sigma_g$  will increase to such large
values that the disk will be unstable to gravitational instabilities (Laughlin
\& Bodenheimer 1994).  The small contribution of the change in the active
layer at the snow line will not be important in these early disks.
Later, the mass accretion-rate decreases and/or the grains grow so that the structure of the MRI-active layer strongly affects the total disk column depth.
In this stage there is a dip in the steady-state surface density profile 
corresponding to the increase in viscosity interior to the snow line.  In 
these disks the pressure gradient is inverted, and solid material accumulates. 
Finally, the mass accretion rate, and therefore surface density, will decline to the point that the entire disk is MRI-active, and this trapping mechanism will no longer be effective.
Regardless of the grain properties this stage will have occured by the time the mass accretion rate drops to the level observed among the weak-line T Tauri stars.

\begin{figure}
	\plotone{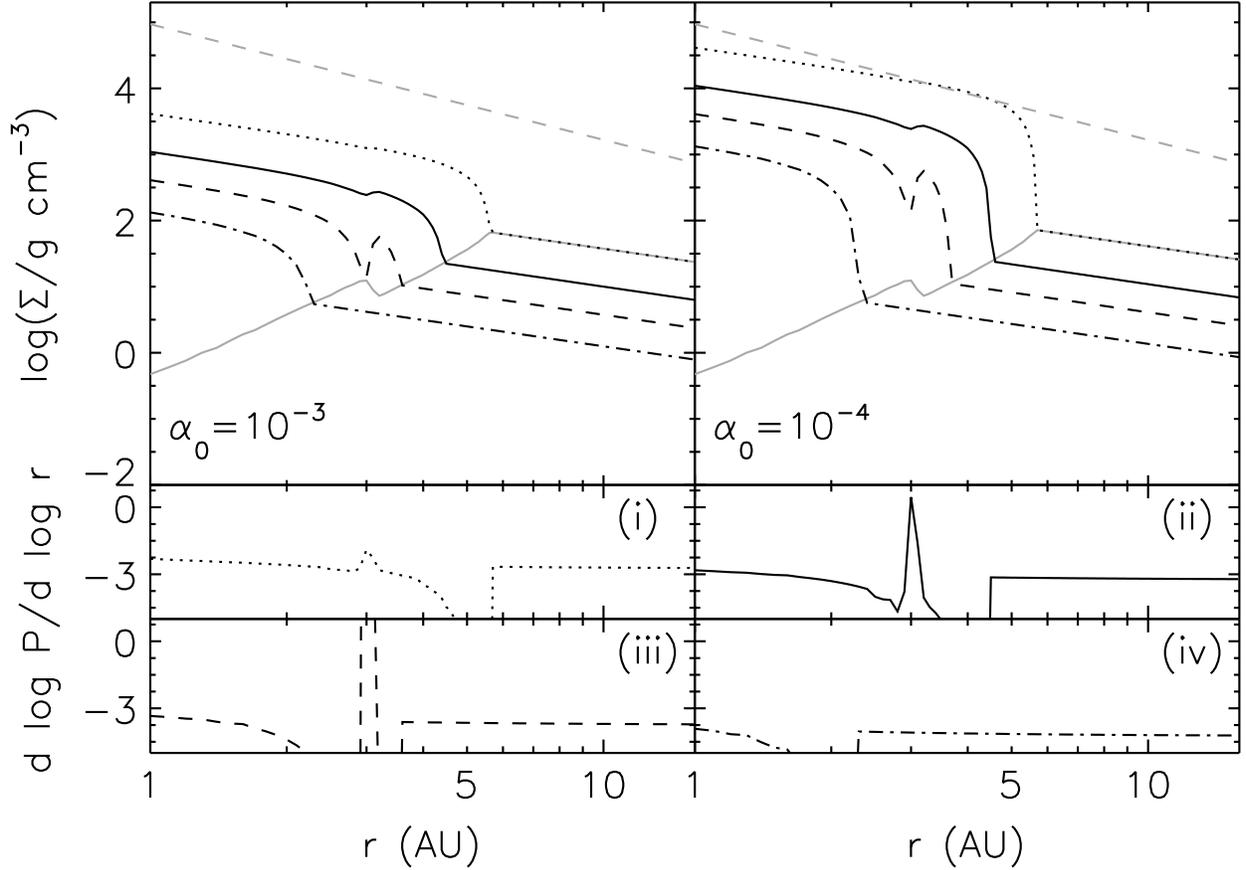}
	\caption{The steady-state profiles for two different alphas and different mass accretion rates $\dot M_g$=$10^{-9}, 3\times10^{-9}, 8\times10^{-9}, 3\times10^{-8}$ $M_{\sun} \rm{year}^{-1}$ (black dot-dashed, dashed, solid and dotted, respectively).  The grey dashed line shows the Toomre stability criteria. 
The grey solid line shows the $\Sigma_A$ with $1 \mu m$ grains with a dust-to-gas ratio changing from 0.005 to 0.011 across the snow line (Lodders 2003). 
Panels (i)-(iv) show the corresponding pressure gradients for each $\dot M_g$, which are independent of $\alpha_0$.}\label{layer}
\end{figure}

To demonstrate these stages 
Fig.~\ref{layer} shows the resulting steady-state surface density profiles for 
a series of accretion rates and illustrative examples of $\alpha_0$ (not intended to constrain the possible range of $\alpha_0$).
We approximate the size of 
the active layer from chemical equilibrium calculations to determine 
the linear stability to MRI turbulence (Ilgner \& Nelson 2006 (model 4); Neal 
Turner private communication).  
In these calculations we assume a grain size of $1 \mu m$. 
If the grain size is substantially smaller than this the column depth of the 
active region at the snow line may be too small to generate turbulence.
The most obvious feature in the surface density profiles is the sharp decrease in the surface density at the edge of the dead-zone (beyond which $\Sigma_g=\Sigma_A$).
The effect that we are interested in is the small change in surface density at the snow line, around 3 AU.  
The curves span the range of phases described abover, so only the middle two have pressure inversions at the snow line.
We note that the assumed temperature profile with $q=0.5$ is not self-consistent with the disk
structure assuming a passively irradiated disk, but as the pressure gradient is only very weakly dependent on the temperature we neglect the self-consistent calculation 
at this time.

\section{Discussion and Conclusions}

We have demonstrated that, in a disk whose evolution is dominated by the
MRI-driven turbulence, the sensitivity of the instability to the grain 
properties creates regions with radial-pressure-gradient inversion and results 
in the accumulation of migrating particles.  
This mechanism provides a potential resolution to the problem of grain retention in protoplanetary disks.
These radial-pressure-gradient inversions occur at the sublimation fronts, 
such as the snow line,  with a modest grain depletion as long as the 
transition occurs over a short radial extent.

In the presence of another viscous mechanism in the MRI-stable zones the disk may be able to reach a quasi-steady-state.
Under these conditions, the solid material will
continue to flow into the pressure maximum until the amount of solid material at 
this radius becomes comparable to the amount of gas in the disk. 
Thereafter, the dust transfers angular momentum to the gas in the
inactive shielded layers, forcing it to rotate at the local Keplerian 
speed and to diffuse to larger radii.  The region of dust accumulation 
then expands slowly outward.  
In the absence of another source of viscosity,   
a steady-state may not be attainable. In this case, gas will
continue to accumulate, maintaining a pressure maximum which will trap solid material until other dust or gas instabilities start to dominate the disk evolution. 

The particle collected at the snow line should rapidly grow through cohesive collisions.  
At the snow line, the grains of all sizes are likely to
be covered with marginally molten frost which can promote the cohesive
nature of the collisions (Bridges {\it et al.} 1996).
The particles may also be able to grow past the ``meter-
sized barrier'' (Weidenschilling 1984;  Cuzzi \& Weidenschilling 2006).
First, as at the pressure maximum all material is moving at Keplarian velocity, large objects will no be ``sand-blasted'' by smaller grains.  If particles grow past the meter-sizescale to the regime where they are not well coupled to the turbulent flow there is no further barrier to their growth as there would be at other poisitions in the disk.
Additionally, as the particles are collected in regions where the midplane is stable to MRI-turbulence, the question of whether meter-sized objects (coupled to the largest eddies) will undergo destructive collisions depends on the properties of whatever mechanism for angular momentum transport may occur in the ``dead zone.''  If the disk is weakly turbulent ($\alpha_0 < 10^{-4}$) or if the trasport of angular momentum is due to a mechansism not involving turbulence then even this aspect of the ``barrier'' can be overcome. 
\acknowledgments
We thank Neal Turner for many enlightening discussions and for providing us with the calculations of the depth of the MRI-active layer.  We also thank our referee, John Chambers, for his helpful comments. This work has been supported 
by NASA (NNG05G-142G, NNG04G-191G, NNG06-GH45G), JPL (1270927), 
NSF(AST-0507424) and IGPP1326.

\clearpage

\clearpage

\end{document}